# A new succinct proof on the equivalence of the Nernst equation and the vanishing heat capacity at absolute zero temperature

Shanhe Su, Jincan Chen[*]

Department of Physics, Xiamen University, Xiamen 361005, People's Republic of China

**Abstract**. By starting from the Euler chain rule of three thermodynamic quantities, it is proved that both the Nernst equation and the vanishing heat capacity at absolute zero temperature are mutually deducible and equivalent. Simultaneously, it is pointed out that the conclusions of the relevant literature are worthy of discussion.

It is important to highlight that several errors have been included in the article [1] published recently in this journal. For example, Eq.(2) in Ref.[1] is the most key equation of the rebuttal article, but essentially it does not contain the Nernst equation; $\lim_{\Delta x \to 0}(\Delta S)_T = 0$ is mistaken for $\lim_{T \to 0}(\Delta S)_T = 0$ and used in the proof process.

The three equations presented in Ref. [1]

$$\left(\frac{\partial S}{\partial T}\right)_x \left(\frac{\partial T}{\partial X}\right)_S \left(\frac{\partial X}{\partial S}\right)_T = -1, \tag{1}$$

$$\left(\frac{\partial S}{\partial T}\right)_x = \lim_{\Delta X \to 0}\frac{(\Delta S)_T}{(\Delta T)_S}, \tag{2}$$

and

$$C_x = T\left(\frac{\partial S}{\partial T}\right)_x = T\frac{(\Delta S)_T}{(\Delta T)_S} \tag{3}$$

clearly demonstrate that $X$ and $x$ represent the same quantity and can be uniformly

---

[*] Email: jcchen@xmu.edu.cn

expressed as $x$ in the following discussion.

It is evident from Eq. (2) and Fig.1 in Ref. [1] that when $\Delta x \to 0$, two curves indicated by $x_1$ and $x_2$ overlap to form a single curve, such that $\lim_{\Delta x \to 0}(\Delta T)_S = 0$ and $\lim_{\Delta x \to 0}(\Delta S)_T = 0$ for any temperature of $T \geq 0$. Obviously, $\lim_{\Delta x \to 0}(\Delta S)_T = 0$ is completely different from the Nernst equation

$$\lim_{T \to 0}(\Delta S)_T = S(0, x_2) - S(0, x_1) = 0. \tag{4}$$

This clearly indicates that Eq. (2) does not contain the Nernst equation and cannot be used to discuss the relationship between $\lim_{T \to 0}(\Delta S)_T = 0$ and $\lim_{T \to 0} C_x = 0$. Recent research [2] has shown that, without any additional assumptions, the Nernst equation can be re-deduced from the experimental data obtained from thermodynamic systems at ultra-low temperatures. Consequently, the physical content represented by the Nernst equation should be referred to as the Nernst statement rather than the Nernst theorem.

It can also be observed from Eqs. (2) and (3) that $\lim_{\Delta x \to 0}$ should be included in the last term of Eq. (3), i.e.,

$$C_x = T\left(\frac{\partial S}{\partial T}\right)_x = \lim_{\Delta x \to 0}[T\frac{(\Delta S)_T}{(\Delta T)_S}]. \tag{5}$$

Similar to Eq. (2), Eq. (5) does not contain the Nernst equation and cannot be used to discuss the relationship between $\lim_{T \to 0}(\Delta S)_T = 0$ and $\lim_{T \to 0} C_x = 0$. However, the author in Ref. [1] asserted that "there is no way to elucidate which of the two in Eq. (2) ---$(\Delta T)_S$ and $(\Delta S)_T$--- vanishes faster as $T \to 0$. With Eq. (3), there is no way to elucidate whether heat capacities vanish or not." It shows clearly that in Ref. [1], $\lim_{\Delta x \to 0}(\Delta S)_T = 0$ was wrongly considered equivalent to $\lim_{T \to 0}(\Delta S)_T = 0$.

Practically, Eq. (1) in Ref. [1] can be directly used to prove that $\lim_{T \to 0}(\Delta S)_T = 0$ and $\lim_{T \to 0} C_x = 0$ are mutually deducible and equivalent [3]. Using the Euler chain rule shown in Eq. (1), one can derive

$$\left(\frac{\partial S}{\partial T}\right)_x \left(\frac{\partial T}{\partial x}\right)_S = -\left(\frac{\partial S}{\partial x}\right)_T. \tag{6}$$

As described in Ref. [1], the condition $\lim_{T \to 0}(\partial S / \partial x)_T$ does not force any further condition on $\lim_{T \to 0}(\partial S / \partial T)_x$ [4,5]. This conclusion is correct, but it cannot be used to negate the fact that $\lim_{T \to 0} C_x = 0$ can be directly derived from $\lim_{T \to 0}(\Delta S)_T = 0$. Now, Eq. (6) is rewritten as

$$C_x \frac{1}{T}\left(\frac{\partial T}{\partial x}\right)_S = -\left(\frac{\partial S}{\partial x}\right)_T. \tag{7}$$

By utilizing the following relations $(\Delta T)_S = \left(\frac{\partial T}{\partial x}\right)_S (\Delta x)_S + \frac{1}{2}\left(\frac{\partial^2 T}{\partial x^2}\right)_S (\Delta x)_S^2 + ...$ and

$(\Delta S)_T = \left(\frac{\partial S}{\partial x}\right)_T (\Delta x)_T + \frac{1}{2}\left(\frac{\partial^2 S}{\partial x^2}\right)_T (\Delta x)_T^2 + ...$, Eq. (7) can be rewritten as

$$\frac{C_x}{T}\{\frac{(\Delta T)_S}{(\Delta x)_S} - [\frac{1}{2}\left(\frac{\partial^2 T}{\partial x^2}\right)_S (\Delta x)_S + ...]\} = -\{\frac{(\Delta S)_T}{(\Delta x)_T} - [\frac{1}{2}\left(\frac{\partial^2 S}{\partial x^2}\right)_T (\Delta x)_T + ...]\}. \tag{8}$$

The high order small quantities on the two sides of Eq. (8) can be expressed as

$$-\{\frac{C_x}{T}[\frac{1}{2}\left(\frac{\partial^2 T}{\partial x^2}\right)_S (\Delta x)_S + ...]\} = [\frac{1}{2}\left(\frac{\partial^2 S}{\partial x^2}\right)_T (\Delta x)_T + ...] + \Delta, \tag{9}$$

where $\Delta$ indicates the difference value produced by the high order small quantities on the two sides of Eq. (8). Compared to the low-order quantity $\frac{(\Delta S)_T}{(\Delta x)_T}$ in Eq. (8), $\Delta$ is negligible, and consequently, Eq. (8) can be simplified to

$$\{\lim_{T \to 0}[-\frac{(\Delta T)_S}{T} C_x]\}\frac{1}{(\Delta x)_S} = [\lim_{T \to 0}(\Delta S)_T]\frac{1}{(\Delta x)_T}. \tag{10}$$

In Eq. (10), $\Delta x$ is determined solely by $x_2$ and $x_1$ and is independent of $T \to 0$. It is clear from Eq. (4) that the Nernst equation does not require $\Delta x = x_2 - x_1$ to approach zero.

In other words, $\Delta x \neq 0$ is a significant component of the Nernst equation. For any adiabatic cooling process starting from an arbitrary low temperature $T$, $-(\Delta T)_S$ is impossibly greater than $T$ and $-(\Delta T)_S/T$ is always a finite value greater than zero [6] but cannot be larger than 1, since +0 K represents the low limit of the absolute temperature [7-9]. Because $\Delta x$ and $-(\Delta T)_S/T$ are finite, Eq. (10) and the Nernst equation $\lim_{T \to 0}(\Delta S)_T = 0$ can be directly used to derive the heat capacity statement [2] $\lim_{T \to 0} C_x = 0$. Additionally, Eq. (10) and the heat capacity statement $\lim_{T \to 0} C_x = 0$ can also be directly used to derive the Nernst equation $\lim_{T \to 0}(\Delta S)_T = 0$. Thus, it has been demonstrated that $\lim_{T \to 0}(\Delta S)_T = 0$ and $\lim_{T \to 0} C_x = 0$ are mutually deducible and equivalent [3,10].

Since $\Delta x$ determined solely by $x_2$ and $x_1$ is independent of whether the process is isothermal or adiabatic, Eq. (10) can be simplified to

$$\lim_{T \to 0}[-\frac{(\Delta T)_S}{T} C_x] = \lim_{T \to 0}(\Delta S)_T. \tag{11}$$

In fact, Eq. (11) can be directly derived from the entropy change $\Delta S = (\Delta S)_x + (\Delta S)_T = 0$ of a reversible adiabatic cooling process [3]. As long as one recognizes that $-(\Delta T)_S/T$ is a finite value greater than zero, Eq. (11) can be conveniently used to demonstrate that $\lim_{T \to 0}(\Delta S)_T = 0$ and $\lim_{T \to 0} C_x = 0$ are mutually deducible and equivalent.

The above proofs are based on the condition that $-(\Delta T)_S/T$ is a finite value greater than zero. Even if this condition is not used, it can be proved that $\lim_{T \to 0}(\Delta S)_T = 0$ and $\lim_{T \to 0} C_x = 0$ are equivalent [10]. For this purpose, Eq. (10) can be rewritten as

$$\{\lim_{T \to 0}[-\frac{(\Delta T)_S}{T}]\}\frac{1}{(\Delta x)_S} = [\lim_{T \to 0}\frac{(\Delta S)_T}{C_x}]\frac{1}{(\Delta x)_T}. \tag{12}$$

Eq.(12) indicates that when $\lim_{T \to 0} C_x = 0$, it is necessary to require $\lim_{T \to 0}(\Delta S)_T = 0$ because $-(\Delta T)_S/T$ cannot be larger than 1. It shows clearly that $\lim_{T \to 0}(\Delta S)_T = 0$ can be derived from Eq. (12) and $\lim_{T \to 0} C_x = 0$. Moreover, according to the Nernst equation and L'Hopital's rule, we have

$$\lim_{T \to 0} S = S_0 = \lim_{T \to 0} \frac{TS}{T} = \lim_{T \to 0} [\frac{\partial (TS)}{\partial T}]_x = \lim_{T \to 0} S + \lim_{T \to 0} (T \frac{\partial S}{\partial T})_x, \quad (13)$$

where $S_0$ is a constant. From Eq. (13), one can obtain

$$\lim_{T \to 0} (T \frac{\partial S}{\partial T})_x = \lim_{T \to 0} C_x = 0. \quad (14)$$

It is proved once again that that $\lim_{T \to 0} (\Delta S)_T = 0$ and $\lim_{T \to 0} C_x = 0$ are mutually deducible and equivalent. By the way, $\lim_{T \to 0} (\Delta S)_T = 0$ and $\lim_{T \to 0} C_x = 0$ are, respectively, two conclusions obtained by extrapolating the simulation results of the experimental data of thermodynamic systems at low temperatures to absolute zero temperature [2]. Although they can be proved to be equivalent, each of them is impossibly deduced from thermodynamic theory.